\documentclass[aps,prb,twocolumn,10pt,superscriptaddress,floatfix,longbibliography]{revtex4-2}
\usepackage{xcolor}

\usepackage{amsmath,amssymb} 
\usepackage{physics}
\usepackage{bm} 
\usepackage [autostyle, english = american]{csquotes}
\MakeOuterQuote{"}
\usepackage{braket}
\usepackage{enumitem}
\usepackage{graphicx} 
\usepackage{comment} 
\usepackage{textcomp} 

\usepackage[english]{babel} 
\usepackage{csquotes}        
\usepackage{graphicx}        
\usepackage{hyperref}        
\usepackage{enumitem}
\setlist{noitemsep,leftmargin=*,topsep=0pt,parsep=0pt}

\usepackage{xcolor} 
\definecolor{lightgray}{gray}{0.6}
\definecolor{medgray}{gray}{0.4}

\usepackage{braket}

\usepackage{hyperref}
\hypersetup{
colorlinks=true,
urlcolor= blue,
citecolor=blue,
linkcolor= blue,
}

\newcommand{\mytitle}{Engineered Dissipation for Thermal-State Tracking Across Quantum Criticality}

\begin{document}

\title{\mytitle}
\author{Saikat Mistry}
\email[]{saikat23@iiserb.ac.in}
\affiliation{Department of Physics, Indian Institute of Science Education and Research, Bhopal, India}

\author{Dibyajyoti Sahu}
\email[]{dibyajyoti20@iiserb.ac.in}
\affiliation{Department of Physics, Indian Institute of Science Education and Research, Bhopal, India}

\author{Suhas Gangadharaiah}
\email[]{suhasg@iiserb.ac.in}
\affiliation{Department of Physics, Indian Institute of Science Education and Research, Bhopal, India}

\date{\today}


\begin{abstract}
Engineering dissipation has emerged as a powerful strategy for controlling quantum many-body dynamics, complementing purely coherent control for state preparation. However, finite-time driving across a quantum critical point generates nonadiabatic excitations and coherences, and existing dissipation-engineering approaches have largely characterized how a fixed environment modifies critical dynamics rather than actively steering a system toward a prescribed thermal trajectory. Here we develop a Markovian dissipative framework for finite-time thermal-state tracking based on engineered population currents with non-negative transition rates, and apply it to the transverse-field Ising model driven across its quantum critical point. Compared with unitary dynamics and a natural thermal bath, the engineered dynamics strongly suppresses thermal excitation deviations and provides additional control over coherence damping through a free common current $Q_k$, while exhibiting a substantially steeper residual scaling, $\delta n_{\mathrm{ex}}^{\mathrm{th}}\sim(v/v_{\mathrm{ch}})^2$, than the Kibble--Zurek behavior recovered for unitary evolution. These results point toward engineered dissipation as a route for controlling thermal-state preparation in driven quantum critical systems.
\end{abstract}

\maketitle

\section{Introduction}

Quantum technologies have undergone remarkable progress over the past decade, stimulating enormous interest in the controlled preparation and manipulation of quantum many-body states. Applications that range from quantum simulation and quantum annealing to adiabatic quantum computation rely heavily on the ability to drive a quantum system from an easily prepared initial Hamiltonian to a target Hamiltonian whose ground or thermal state encodes the solution to a computational or physical problem~\cite{Farhi2001,Albash2018,Georgescu2014,laumann2015quantum, albash2015reexamination}. Consequently, finite-time quantum control protocols have become one of the central themes in modern quantum information science, condensed matter physics, and quantum thermodynamics.

For realistic implementations, however, no quantum system is perfectly isolated. Coupling to the surrounding environment induces decoherence, dissipation, and thermalization, making the dynamics intrinsically open~\cite{breuer2002theory,Rivas2012}. Rather than being solely viewed as a detrimental effect, environmental interactions have increasingly been recognized as a valuable resource for quantum state engineering when employed under a controlled setup~\cite{Kraus2008,Diehl2008,verstraete2009quantum,Poyatos1996}. Recent advances have demonstrated that properly engineered dissipation can stabilize quantum phases, prepare entangled states~\cite{harrington2022engineered}, realize autonomous quantum error correction, and even accelerate thermalization protocols~\cite{alipour2020shortcuts}.

The preparation of equilibrium states through the continuous variation of a transverse parameter forms the basis of numerous quantum technologies. During an adiabatic protocol, a control parameter is slowly varied such that the instantaneous state continuously follows the desired eigenstate of the Hamiltonian. Similar protocols also underlie adiabatic annealing and quantum adiabatic optimization, where computational problems are encoded in the final Hamiltonian while the system is initialized in the ground state of a simple reference Hamiltonian.

The requirement of adiabaticity, however, poses a severe practical limitation. According to the adiabatic theorem, the driving time must be sufficiently long compared to the characteristic timescale set by the minimum spectral gap~\cite{Albash2018,Kato1950,jansen2007bounds}. In many-body systems, as a quantum critical point is approached, the excitation gap closes in the thermodynamic limit, causing the adiabatic timescale to diverge~\cite{Sachdev2011,Dziarmaga2010}. Consequently, finite-rate driving inevitably generates nonadiabatic excitations and defects whose dependence on the driving rate is described by the quantum Kibble--Zurek mechanism~\cite{Kibble1976,Zurek1985,Zurek2005,Polkovnikov2005,Dziarmaga2005,schutzhold2006sweeping,uhlmann2007vortex}, preventing perfect state preparation and substantially reducing the efficiency of quantum annealing and adiabatic quantum computation.

Considerable effort has been devoted to suppressing such finite-rate excitations through shortcuts to adiabaticity \cite{Torrontegui2013}, particularly counterdiabatic driving~\cite{DemirplakRice2003,Berry2009}. 
Practical constructions of the corresponding adiabatic gauge potential include variational \cite{Sels2017} and Krylov-space approaches \cite{takahashi2024shortcuts}, while counterdiabatic protocols have also recently been demonstrated experimentally in driven many-body systems \cite{visuri2026digitized}. Optimal-control strategies provide another route to mitigating finite-rate excitations~\cite{caneva2009optimal}. Many of these
approaches primarily concern coherent control, although extensions of
transitionless driving to Lindblad open-system dynamics have also been
developed~\cite{Vacanti2014}. 
Related STA constructions have also been developed for prescribed thermal-state tracking in open quantum critical systems, including the transverse-field Ising model~\cite{mahunta2025shortcuts}, with recent extensions to long-range interacting critical systems \cite{mahunta2026long}. The interplay between finite-rate critical dynamics and environmental dissipation has also been extensively investigated in open many-body systems, where thermal relaxation and dissipative timescales introduce additional excitation mechanisms and scaling regimes that can modify the conventional closed-system Kibble--Zurek behavior~\cite{Patane2008,Nalbach2015,RossiniVicari2020}. Related open-system
control protocols have further been developed to accelerate relaxation
toward equilibrium states~\cite{Dann2019}. 
The idea of engineering a dissipative channel whose net population current reproduces a prescribed thermal trajectory has itself been introduced in a general two-level setting~\cite{alipour2020shortcuts}, but the construction of explicit non-negative transition rates ensuring a physically valid Markovian generator, and its extension to a many-body critical setting, has not been addressed. In this setting, the problem of following a prescribed instantaneous thermal state involves not only controlling nonadiabatic excitation production, but also generating the population redistribution required by the changing thermal target. Existing studies have largely characterized how a given environment modifies the critical dynamics; the complementary problem of systematically engineering Markovian dissipative transition rates to follow a prescribed finite-temperature trajectory during finite-time driving across a quantum critical region remains comparatively less explored.

Motivated by this problem, and building on the trajectory-based open-system framework of ~\cite{alipour2020shortcuts}, we investigate whether the population redistribution required by a prescribed instantaneous Gibbs trajectory can be realized within a strictly Markovian GKSL description. We construct a family of non-negative transition rates whose net population current reproduces the prescribed thermal-population derivative, while retaining freedom to control coherence damping independently of the required net transfer. We then apply this construction to finite-time driving across the quantum critical region of the transverse-field Ising model and assess its ability to suppress deviations from the target thermal state. We apply the framework to the transverse-field Ising model under linear ramps across its quantum critical point and compare the engineered dynamics with both unitary evolution and coupling to a conventional thermal reservoir. We characterize the resulting thermal excitation deviation, coherence dynamics, and finite-rate scaling, and further examine the protocol for a ramp traversing both critical points. The transverse-field Ising model thereby provides a concrete many-body setting for assessing how engineered dissipation can control finite-time thermal-state tracking through quantum critical regions.

The remainder of the paper is organized as follows. In Sec.~\ref{sec:theory}, we
develop the general theoretical framework for prescribed thermal-state
tracking, beginning with the instantaneous Gibbs trajectory and
Markovian open-system dynamics, followed by the natural thermal-bath
reference and the engineered population-current construction. In
Sec.~\ref{sec:tfim}, we specialize the framework to the transverse-field Ising
model, introduce the finite-time driving protocols across its quantum
critical points, and formulate the corresponding mode-resolved
thermalization dynamics. Sec.~\ref{sec:results} presents the numerical results,
including the excited-state population dynamics, thermal excitation
deviation, suppression of dynamically generated coherence, and the
finite-rate scaling of the residual thermal deviation. Finally,
Sec.~\ref{sec:inference} summarizes the main results, discusses considerations for
experimental implementation, and outlines directions for further
work.

\section{Theoretical Framework\label{sec:theory}}
\subsection{Prescribed Thermal Trajectory\label{subsec:prescribe}}

We begin by defining the target mixed-state trajectory that the driven open-system dynamics is intended to follow. Throughout this work, we set $\hbar=k_{\mathrm B}=1$. In the
numerical analysis of the transverse-field Ising model, we additionally
set the Ising coupling $J=1$, which fixes the energy scale.
At each instant, the desired state is taken to be the instantaneous Gibbs state associated with the
system Hamiltonian $H_S(t)$~\cite{Breuer2002},
\begin{equation}
    \rho_{\star}(t)
    =
    \frac{e^{-\beta H_S(t)}}{Z(t)}
    =
    \sum_{n=1}^{r}
    \lambda_n(t)
    |n_t\rangle\langle n_t|,
    \label{eq:target_thermal_state}
\end{equation}
where $Z(t)=\mathrm{Tr}[e^{-\beta H_S(t)}]$ is the instantaneous partition function, $\{|n_t\rangle\}$ are the instantaneous eigenstates of $H_S(t)$, and $\lambda_n(t)$ are the corresponding thermal occupation probabilities satisfying $\lambda_n(t)\geq0$ and $\sum_n\lambda_n(t)=1$.
Hereafter, $\rho_{\star}(t)$ denotes the prescribed thermal target, while $\rho(t)$ denotes the state generated by the actual finite-time dynamics; in general, the two will not coincide, since finite-rate driving will generate deviations of the instantaneous populations from $\lambda_n(t)$ and also off-diagonal coherences in the instantaneous energy eigenbasis.
These two contributions provide complementary measures of the departure of $\rho(t)$ from the prescribed thermal trajectory.

Since purely unitary evolution preserves the spectrum of the density
operator, it cannot by itself generate the time-dependent redistribution
of occupation probabilities required by the prescribed Gibbs state in
Eq.~\eqref{eq:target_thermal_state}. Therefore, following a changing thermal trajectory necessarily
requires a nonunitary contribution to the dynamics. Following the trajectory-based approach to shortcuts to adiabaticity introduced by Alipour \textit{et al.}~\cite{alipour2020shortcuts} (see also Ref.~\cite{dupays2021shortcuts}), which targets the eigenbasis of the density operator itself rather than that of the instantaneous Hamiltonian as in conventional counterdiabatic driving~\cite{DemirplakRice2003,Berry2009}, the spectral
decomposition of the prescribed state separates its evolution into the
rotation of the instantaneous eigenbasis and the redistribution of its
occupation probabilities:
\begin{equation}
    \partial_t \rho_{\star}(t)
    =
    -i\left[H_1(t),\rho_{\star}(t)\right]
    +
    \sum_n
    \dot{\lambda}_n(t)
    |n_t\rangle\langle n_t|.
    \label{eq:target_state_derivative}
\end{equation}
The operator
\begin{equation}
    H_1(t)
    =
    i\sum_n
    \left[
    |\partial_t n_t\rangle\langle n_t|
    -
    \langle n_t|\partial_t n_t\rangle
    |n_t\rangle\langle n_t|
    \right]
    \equiv H_{\mathrm{CD}}(t)
    \label{eq:geometric_hamiltonian}
\end{equation}
is the geometric contribution responsible for transporting the instantaneous eigenbasis; it corresponds to the counterdiabatic Hamiltonian of transitionless quantum driving~\cite{DemirplakRice2003,Berry2009}, here generalized to the eigenbasis of $\rho_\star(t)$.
The second term in Eq.~\eqref{eq:target_state_derivative}, in contrast, describes the population redistribution required for the state to remain on the prescribed thermal manifold. This population contribution, identified in Ref.~\cite{alipour2020shortcuts} as the quantity that a dissipative channel must reproduce, is the central quantity relevant to the dissipative construction developed
below.
Since coherent evolution cannot change the eigenvalues of the density operator, the nonunitary dynamics must generate transition currents whose net effect reproduces the prescribed population derivatives $\dot{\lambda}_n(t)$.
Importantly, specifying this net population transfer does not uniquely
determine the individual transition processes; as we show in the
following subsections, this freedom permits a family of dissipative
realizations with non-negative transition rates that reproduce the
same prescribed target-population derivative.

\subsection{Markovian Open-System Dynamics\label{subsec:open}}

Throughout this work, we restrict the dynamics to the weak-coupling Markovian regime, so that the reduced state of the system is described by the Gorini-Kossakowski-Sudarshan-Lindblad (GKSL) master equation ~\cite{gorini1976completely,lindblad1976generators},
\begin{equation}
    \dot{\rho}(t)
    =
    -i\left[H_S(t),\rho(t)\right]
    +
    \mathcal{D}[\rho(t)],
    \label{eq:gksl_master}
\end{equation}
where $H_S(t)$ is the time-dependent system Hamiltonian and $\mathcal{D}$ denotes the dissipative contribution. Unlike the prescribed Gibbs state, which is diagonal in the instantaneous energy eigenbasis, $\rho(t)$ by construction is not constrained to remain diagonal during the evolution. Consequently, in general $[H_S(t),\rho(t)]\neq0$, and finite-rate driving will generate coherences between instantaneous energy eigenstates in addition to deviations from the prescribed thermal populations.

For a Markovian evolution, the dissipative contribution takes the standard GKSL form~\cite{gorini1976completely, lindblad1976generators,breuer2002theory}
\begin{equation}
    \mathcal{D}[\rho]
    =
    \sum_{\alpha} \kappa_{\alpha}
    \left(
        A_{\alpha}\rho A_{\alpha}^{\dagger}
        -
        \frac{1}{2}
        \left\{
            A_{\alpha}^{\dagger}A_{\alpha},\rho
        \right\}
    \right),
    \label{eq:gksl_dissipator}
\end{equation}
where $A_{\alpha}$ are the Lindblad jump operators describing the allowed dissipative processes and $k_{\alpha}$ are their corresponding transition rates. For $k_{\alpha}\geq0$, the GKSL generator defines a completely positive and trace-preserving (CPTP) dynamical evolution, so that an initially physical density operator remains Hermitian, normalized, and positive semidefinite throughout the evolution ~\cite{rivas2014quantum}. The objective of the construction developed in following subsections is therefore to design dissipative transition channels whose population currents drive $\rho(t)$ toward the prescribed thermal populations while maintaining non-negative transition rates and, consequently, the standard Markovian GKSL structure.

\subsection{Natural Thermal-Bath Dynamics\label{subsec:natural_bath}}

Before constructing the engineered dissipative channels that realize this objective, we first consider the conventional dissipative dynamics generated by a bosonic thermal reservoir. This natural-bath dynamics provides the reference against which the engineered protocol introduced subsequently is compared. The total Hamiltonian of the system and environment is written as
\begin{equation}
    H_{\mathrm{tot}}(t)
    =
    H_S(t)+H_B+H_{SB},
    \label{eq:total_hamiltonian_bath}
\end{equation}
where $H_S(t)$ denotes the driven system Hamiltonian, $H_B$ describes the bosonic thermal reservoir, and $H_{SB}$ represents the system--bath interaction. Within the weak-coupling Born-Markov and secular approximations, the reduced system dynamics takes the GKSL form introduced in the
previous subsection~\cite{Davies1974,Breuer2002}, with the dissipative transition rates determined by the spectral properties and temperature of the reservoir.

We consider an Ohmic reservoir with an exponential high-frequency cutoff ~\cite{Leggett1987,Weiss2012},
\begin{equation}
    S(\omega)
    =
    2\pi\eta\,\omega
    e^{-\omega/\omega_c},
    \label{eq:ohmic_spectral_density}
\end{equation}
where $ \eta$ denotes the system-bath coupling strength and $\omega_c$ is the cutoff frequency ~\cite{caldeira1983quantum}. At an inverse bath temperature $\beta_B=1/T_B$, the corresponding Bose--Einstein occupation factor is
\begin{equation}
    n_B(\omega)
    =
    \frac{1}{e^{\beta_B\omega}-1}.
    \label{eq:bose_distribution}
\end{equation}
The bath-induced excitation and relaxation rates are then given as ~\cite{Breuer2002,Alicki1976}
\begin{align}
    \kappa_{\uparrow}(\omega)
    &=
    S(\omega)n_B(\omega),
    \label{eq:natural_excitation_rate}
    \\
    \kappa_{\downarrow}(\omega)
    &=
    S(\omega)\left[n_B(\omega)+1\right],
    \label{eq:natural_relaxation_rate}
\end{align}
and satisfy the thermal detailed-balance condition~\cite{Alicki1976,Spohn1978},
\begin{equation}
    \frac{\kappa_{\uparrow}(\omega)}
         {\kappa_{\downarrow}(\omega)}
    =
    e^{-\beta_B\omega}.
    \label{eq:detailed_balance}
\end{equation}

For a driven two-level sector with instantaneous ground and excited states $|g_t\rangle$ and $|e_t\rangle$, respectively, the natural thermal reservoir induces excitation and relaxation through the instantaneous jump operators
\begin{equation}
    L_{\uparrow}(t)
    =
    |e_t\rangle\langle g_t|,
    \qquad
    L_{\downarrow}(t)
    =
    |g_t\rangle\langle e_t|,
    \label{eq:natural_jump_operators}
\end{equation}
with the corresponding transition frequency set by the instantaneous energy separation,
\begin{equation}
    \omega(t)
    =
    E_e(t)-E_g(t).
    \label{eq:instantaneous_transition_frequency}
\end{equation}
The natural thermal-bath dissipator therefore takes the form
\begin{align}
    \mathcal{D}_{\mathrm{Nat}}[\rho]
    ={}&
    \kappa_{\uparrow}[\omega(t)]
    \left(
        L_{\uparrow}\rho L_{\uparrow}^{\dagger}
        -
        \frac{1}{2}
        \left\{
            L_{\uparrow}^{\dagger}L_{\uparrow},\rho
        \right\}
    \right)
    \nonumber\\
    &+
    \kappa_{\downarrow}[\omega(t)]
    \left(
        L_{\downarrow}\rho L_{\downarrow}^{\dagger}
        -
        \frac{1}{2}
        \left\{
            L_{\downarrow}^{\dagger}L_{\downarrow},\rho
        \right\}
    \right).
    \label{eq:natural_dissipator}
\end{align}
The natural reservoir therefore induces both thermal excitation and relaxation, with their relative strengths determined by the bath temperature through detailed balance. In the numerical analysis below, this conventional thermal-bath dynamics serves as the dissipative reference against which the engineered population-transfer protocol is compared.

\subsection{Engineered Population-Current Dynamics \label{subsec:engineer}}

As discussed in Subsec.~\ref{subsec:prescribe}, the net population current required to reproduce a prescribed thermal trajectory, when evaluated on the target populations, was identified by Alipour \textit{et al.}~\cite{alipour2020shortcuts}. This condition alone, however, fixes only the net current and leaves the individual excitation and relaxation rates entering the dissipator undetermined; a generic choice satisfying the net-current condition need not yield non-negative rates, and can therefore fall outside the standard Markovian GKSL structure. We now construct, from this net-current condition an explicit family of transition rates $W_{10}(t),W_{01}(t)\geq0$ that reproduces the required population current while remaining within a physically valid, completely positive and trace-preserving Markovian generator, and identify the minimal such realization.

In the instantaneous energy eigenbasis $\{|0(t)\rangle,|1(t)\rangle\}$ of $H_S(t)$, the state of the system is written as
\begin{equation}
    \rho(t)=
    \begin{pmatrix}
        \rho_{00}(t) & \rho_{01}(t)\\
        \rho_{10}(t) & \rho_{11}(t)
    \end{pmatrix},
    \label{eq:rho_instantaneous_basis}
\end{equation}
where the diagonal elements represent the instantaneous populations, while the off-diagonal elements describe coherences between the two instantaneous energy eigenstates.

To engineer population transfer between these levels, we introduce the instantaneous transition operators, following the trajectory-based open-system constructions of \cite{alipour2020shortcuts, mahunta2025shortcuts}
\begin{equation}
    L_{ij}(t)
    =
    |i(t)\rangle\langle j(t)|,
    \qquad
    i,j\in\{0,1\},\quad i\neq j,
    \label{eq:eng_jump_operator}
\end{equation}
in terms of which the corresponding engineered dissipator is
\begin{equation}
    \mathcal{D}^{\mathrm{eng}}[\rho]
    =
    \sum_{i\neq j}
    W_{ij}(t)
    \left(
        L_{ij}\rho L_{ij}^{\dagger}
        -
        \frac{1}{2}
        \left\{
            L_{ij}^{\dagger}L_{ij},
            \rho
        \right\}
    \right),
    \label{eq:eng_dissipator}
\end{equation}
where $W_{ij}(t)$ is the transition rate from state $j$ to state $i$. Evaluating Eq.~\eqref{eq:eng_dissipator} in the instantaneous energy basis, the matrix elements take the following compact form
\begin{equation}
\mathcal{D}^{\mathrm{eng}}[\rho]
=
\begin{pmatrix}
W_{01}\rho_{11}-W_{10}\rho_{00}
&
-\dfrac{W_{01}+W_{10}}{2}\rho_{01}
\\[8pt]
-\dfrac{W_{01}+W_{10}}{2}\rho_{10}
&
W_{10}\rho_{00}-W_{01}\rho_{11}
\end{pmatrix}.
\label{eq:eng_dissipator_matrix}
\end{equation}
The diagonal contribution therefore generates population transfer between the instantaneous energy levels, while the off-diagonal contribution damps the corresponding coherence at a rate set by the total transition activity $W_{01}+W_{10}$; this distinction will later allow us to characterize both population tracking and coherence suppression under the engineered dynamics.

The prescribed instantaneous thermal state introduced in Subsec.~\ref{subsec:prescribe} takes the form
\begin{equation}
    \rho_{\star}(t)
    =
    \lambda_{0}(t)
    |0(t)\rangle\langle0(t)|
    +
    \lambda_{1}(t)
    |1(t)\rangle\langle1(t)|,
    \label{eq:target_state}
\end{equation}
with $\lambda_{0}(t)+\lambda_{1}(t)=1$. The engineered transition rates are chosen by imposing the required population current on the prescribed thermal state. For the target excited-state population, this design condition reads
\begin{equation}
    \dot{\lambda}_{1}(t)
    =
    W_{10}(t)\lambda_{0}(t)
    -
    W_{01}(t)\lambda_{1}(t).
    \label{eq:target_population_constraint}
    \end{equation}
Equation~(\ref{eq:target_population_constraint}) is imposed on the prescribed target populations $\lambda_i(t)$. During the actual evolution, the same rates act on the dynamical populations $p_i(t)$, for which the corresponding dissipative current is $W_{10}(t)p_0(t)-W_{01}(t)p_1(t)$; it therefore coincides with $\dot{\lambda}_1(t)$ only when the actual populations coincide with their target values.
We define the associated excitation and relaxation probability currents as
\begin{equation}
    \mathcal{I}_{10}(t)
    =
    W_{10}(t)\lambda_{0}(t),
    \qquad
    \mathcal{I}_{01}(t)
    =
    W_{01}(t)\lambda_{1}(t),
    \label{eq:population_currents}
\end{equation}
so that the required population dynamics reduces to
\begin{equation}
    \dot{\lambda}_{1}(t)
    =
    \mathcal{I}_{10}(t)-\mathcal{I}_{01}(t).
    \label{eq:net_population_current}
\end{equation}
Eq.~\eqref{eq:net_population_current} constrains only the net population current and therefore does not uniquely determine the two individual transition currents. Introducing the positive-part function $[x]_+\equiv\max(x,0)$, the general non-negative current decomposition can be written as
\begin{align}
    \mathcal{I}_{10}(t)
    &=
    [\dot{\lambda}_{1}(t)]_+
    +Q(t),
    \label{eq:J10_general}
    \\
    \mathcal{I}_{01}(t)
    &=
    [-\dot{\lambda}_{1}(t)]_+
    +Q(t),
    \label{eq:J01_general}
\end{align}
where $Q(t)\geq0$ is an arbitrary common population current. Since $[x]_+-[-x]_+=x$, the contribution $Q(t)$ cancels identically from the net current in Eq.~\eqref{eq:net_population_current} and therefore does not alter the prescribed population derivative.

The corresponding family of engineered transition rates is consequently
\begin{equation}
\begin{aligned}
W_{10}(t)
&=
\frac{
[\dot{\lambda}_{1}(t)]_{+}+Q(t)
}{
\lambda_{0}(t)
},
\\[4pt]
W_{01}(t)
&=
\frac{
[-\dot{\lambda}_{1}(t)]_{+}+Q(t)
}{
\lambda_{1}(t)
}.
\end{aligned}
\label{eq:general_engineered_rates}
\end{equation}
For a finite-temperature target state, $\lambda_{0}(t),\lambda_{1}(t)>0$, and together with $Q(t)\geq0$, Eq.~\eqref{eq:general_engineered_rates} guarantees $W_{10}(t)\geq0$ and $W_{01}(t)\geq0$; the engineered generator therefore retains the standard Markovian GKSL form for any non-negative choice of $Q(t)$. This freedom in $Q(t)$ represents a family of excitation and relaxation processes that produce the same prescribed net target-population current.

For the numerical implementation considered below, we choose the minimal-current realization $Q(t)=0$, which yields
\begin{equation}
    W_{10}(t)
    =
    \frac{[\dot{\lambda}_{1}(t)]_+}
         {\lambda_{0}(t)},
    \qquad
    W_{01}(t)
    =
    \frac{[-\dot{\lambda}_{1}(t)]_+}
         {\lambda_{1}(t)}
    .
    \label{eq:minimal_engineered_rates}
\end{equation}
Thus, an increasing target excited-state population activates only the excitation current, while a decreasing target population activates only the relaxation current: the choice $Q=0$ provides the minimal non-negative transition-current realization of the prescribed thermal population dynamics.

\section{Application to the Transverse-Field Ising Model}
\label{sec:tfim}

The engineered population-current formalism developed in Subsec.~\ref{subsec:engineer} is formulated for a general driven two-level system and is therefore applicable to any qubit-type system whose instantaneous dynamics reduces to an effective two-level description, a structure of particular relevance to quantum annealing. Here we illustrate the formalism using the one-dimensional transverse-field Ising model (TFIM), since, through the Jordan--Wigner and Fourier transformations, this model maps onto independent Bogoliubov--de Gennes (BdG) sectors, each realizing precisely such an effective two-level system. This decomposition provides the mode-resolved structure required for implementing the engineered dissipative dynamics developed above.

\subsection{Transverse-Field Ising Hamiltonian}
\label{subsec:tfim_hamiltonian}

We consider the one-dimensional transverse-field Ising model described by~\cite{Pfeuty1970,Sachdev2011}
\begin{equation}
    H_S(t)
    =
    -J\sum_{j=1}^{L}
    \sigma_j^x\sigma_{j+1}^x
    -
    g(t)\sum_{j=1}^{L}
    \sigma_j^z ,
    \label{eq:tfim_hamiltonian}
\end{equation}
where $J$ denotes the nearest-neighbor Ising interaction strength and $g(t)$ is the time-dependent transverse-field strength; both $J$ and $g(t)$ set energy scales of the system.

Applying the Jordan--Wigner transformation ~\cite{JordanWigner1928,Pfeuty1970} maps the spin Hamiltonian onto a quadratic fermionic model, and a subsequent Fourier transformation decomposes it into dynamically independent momentum sectors,
\begin{equation}
    H_S(t)
    =
    \sum_{k>0}
    \Psi_k^\dagger H_k(t)\Psi_k ,
    \label{eq:tfim_k_decomposition}
\end{equation}
where
\begin{equation}
    \Psi_k
    =
    \begin{pmatrix}
        c_k\\
        c_{-k}^{\dagger}
    \end{pmatrix}
    \label{eq:nambu_spinor}
\end{equation}
is the Nambu spinor, and the corresponding BdG Hamiltonian can be written as
\begin{equation}
        H_k(t)
        =
        2
        \left[
            \bigl(g(t)-J\cos k\bigr)\sigma_z
            +
            J\sin k\,\sigma_x
        \right].
    \label{eq:tfim_bdg}
\end{equation}
For an even chain of length $L$, we work in the even-parity sector
with antiperiodic boundary conditions, for which
\begin{equation}
k_m=\frac{(2m+1)\pi}{L},
\quad
m=0,\ldots,\frac{L}{2}-1,
\quad
N_k=\frac{L}{2}.
\end{equation}
Each momentum sector therefore constitutes an independent effective two-level system, evolving under its own $H_k(t)$ ~\cite{Dziarmaga2005,Dziarmaga2010}; since the momentum sectors are dynamically independent in this way, the general two-level engineered-dissipation formalism of Sec.~\ref{sec:theory} applies directly and separately to each mode $k$, with every population, current, and rate introduced there ($\rho\to\rho_k$, $\lambda_{0,1}\to\lambda_{0,1,k}$, $W_{ij}\to W_{ij,k}$) carried over mode-by-mode without further modification.

The instantaneous eigenenergies of $H_k(t)$ are
\begin{equation}
    E_{\pm,k}(t)=\pm\epsilon_k(t),
\end{equation}
where the positive quasiparticle energy is
\begin{equation}
    \epsilon_k(t)
    =
    2
    \sqrt{
        \bigl[g(t)-J\cos k\bigr]^2
        +
        J^2\sin^2 k
    }.
    \label{eq:tfim_dispersion}
\end{equation}
The resulting quasiparticle dispersion is the standard spectrum of
the one-dimensional transverse-field Ising chain
~\cite{Pfeuty1970,Sachdev2011}. Denoting the corresponding instantaneous ground and excited states by $|g_k(t)\rangle$ and $|e_k(t)\rangle$, respectively, the instantaneous two-level energy separation is therefore
\begin{equation}
    \Delta_k(t)
    =
    E_{e,k}(t)-E_{g,k}(t)
    =
    2\epsilon_k(t).
    \label{eq:tfim_level_spacing}
\end{equation}

In the thermodynamic limit, the quasiparticle gap closes at ~\cite{Pfeuty1970,Sachdev2011}
\begin{equation}
    g_c=\pm J,
    \label{eq:tfim_critical_points}
\end{equation}
separating the ferromagnetic phase, $|g|<J$, from the paramagnetic phase, $|g|>J$. In the thermodynamic limit at zero temperature, the ferromagnetic phase is characterized by a nonvanishing spontaneous magnetization
along the Ising interaction direction~\cite{Pfeuty1970},
\begin{equation}
    m_x
    =
    \begin{cases}
    \left[
        1-\left(\dfrac{g}{J}\right)^2
    \right]^{1/8},
    & |g|<J,\\[6pt]
    0,
    & |g|\geq J.
    \end{cases}
    \label{eq:tfim_magnetization}
\end{equation}
The vanishing of $m_x$ at $|g|=J$, together with the closing of the bulk excitation gap, identifies the quantum critical points separating the two phases.

\begin{figure}[t]
    \centering
    \includegraphics[width=\columnwidth]
    {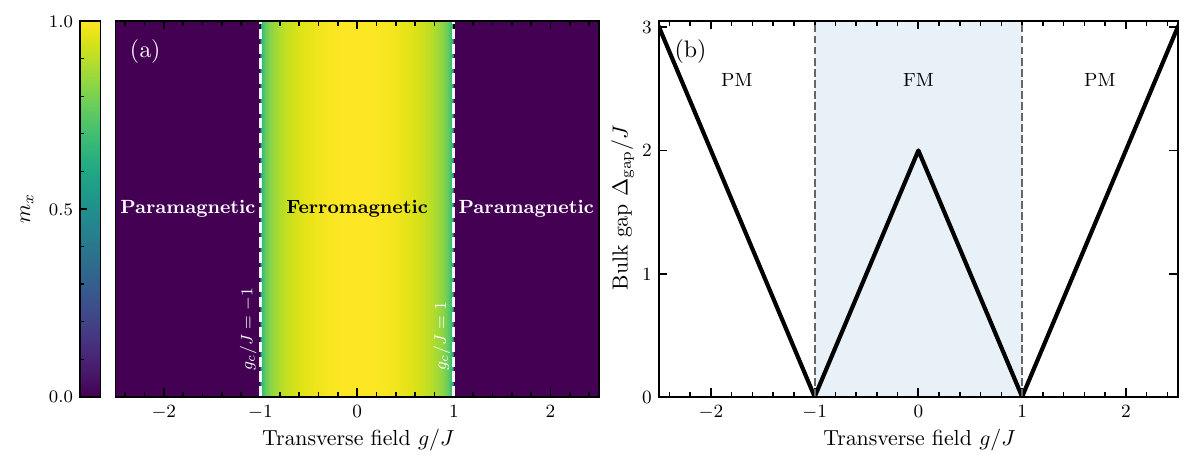}
    \caption{
    Equilibrium phase structure of the transverse-field Ising model.
    (a) Spontaneous magnetization $m_x$, distinguishing the
    ferromagnetic ($|g/J|<1$) and paramagnetic ($|g/J|>1$) phases.
    (b) Bulk quasiparticle excitation gap, which closes at the quantum
    critical points $g/J=\pm1$.
    }
    \label{fig:tfim_phase}
\end{figure}

Fig.~\ref{fig:tfim_phase} summarizes the equilibrium phase structure of the model: the spontaneous magnetization distinguishes the ferromagnetic region from the surrounding paramagnetic phases, while the bulk quasiparticle gap closes at $g=\pm J$. These equilibrium properties identify the critical regions traversed by the finite-time driving protocols considered below.

Throughout the following analysis, energies are expressed in units of the Ising coupling by setting $J=1$; the transverse-field strength $g(t)$ is therefore expressed in units of $J$, and the quantum critical points are located at $g_c=\pm1$.

\subsection{Finite-Time Driving Across the Quantum Critical Point}
\label{subsec:tfim_driving}

To investigate the nonequilibrium dynamics across the quantum critical region, the transverse field is varied linearly in time according to
\begin{equation}
    g(t)
    =
    g_i
    +
    \frac{g_f-g_i}{\tau}\,t,
    \qquad
    0\leq t\leq\tau ,
    \label{eq:linear_ramp}
\end{equation}
where $g_i$ and $g_f$ denote the initial and final transverse-field strengths and $\tau$ is the total duration of the protocol. The corresponding ramp velocity is defined as
\begin{equation}
    v
    \equiv
    |\dot{g}|
    =
    \frac{|g_f-g_i|}{\tau}.
    \label{eq:ramp_velocity}
\end{equation}
The primary protocol considered in this work drives the system across the critical point $g_c=1$, from the paramagnetic phase into the ferromagnetic phase. Since the long-wavelength quasiparticle gap closes in the thermodynamic limit as the critical point is approached, the characteristic relaxation timescale diverges ~\cite{Zurek2005,Polkovnikov2005,Dziarmaga2005,Dziarmaga2010}; for a finite-rate ramp, the evolution consequently departs from the adiabatic limit in the vicinity of the critical point, generating nonadiabatic excitations and coherences in the instantaneous energy eigenbasis.

Within the open-system setting considered here, these finite-time effects cause the actual dynamical state $\rho_k(t)$ to depart from the prescribed instantaneous thermal target $\rho_{\star,k}(t)$ introduced in Sec.~\ref{sec:theory}. The critical crossing therefore provides a stringent test of the engineered population-current dynamics, particularly their ability to keep the dynamical populations close
to the prescribed thermal populations while suppressing coherences generated during the driven evolution. 

In addition to the single-critical-point protocol, we consider an extended ramp traversing both $g_c=1$ and $g_c=-1$, driving the system successively through the paramagnetic, ferromagnetic, and opposite paramagnetic regimes; this provides a robustness test of the engineered dynamics under successive critical crossings.


\subsection{Mode-Resolved Thermalization Protocol}
\label{subsec:tfim_thermalization}

Using the mode decomposition and quasiparticle spectrum introduced in
Eqs.~\eqref{eq:tfim_k_decomposition} and \eqref{eq:tfim_bdg}, we now specialize the population-current construction
of Subsec.~\ref{subsec:engineer} to the TFIM momentum sectors.

For each $k$ sector, the prescribed thermal target is the instantaneous
Gibbs state
\begin{equation}
\rho_{\star,k}(t)
=
\frac{e^{-\beta H_k(t)}}{Z_k(t)},
\end{equation}
where
\begin{equation}
Z_k(t)
=
\mathrm{Tr}\!\left[e^{-\beta H_k(t)}\right]
=
2\cosh\!\left[\beta\epsilon_k(t)\right].
\end{equation}

Using the instantaneous eigenenergies given in Eq.~\eqref{eq:tfim_dispersion}, the
corresponding ground- and excited-state thermal populations are
\begin{equation}
\lambda_{0,k}(t)
=
\frac{1}{1+e^{-\beta\Delta_k(t)}},
\qquad
\lambda_{1,k}(t)
=
\frac{1}{1+e^{+\beta\Delta_k(t)}},
\end{equation}

Since the transverse field varies during the protocol, the
instantaneous quasiparticle spectrum, and hence the prescribed thermal
populations, evolve continuously. The excited-state thermal
population therefore obeys
\begin{equation}
\dot{\lambda}_{1,k}(t)
=
-\beta\lambda_{0,k}(t)\lambda_{1,k}(t)
\dot{\Delta}_k(t).
\end{equation}

For the linear ramp defined in Eq.~\eqref{eq:linear_ramp}, the corresponding rate of
change of the quasiparticle energy is
\begin{equation}
\dot{\epsilon}_k(t)
=
\frac{
2\,[g(t)-J\cos k]\,\dot{g}(t)
}{
\sqrt{
[g(t)-J\cos k]^2+J^2\sin^2 k
}
}.
\end{equation}

The required directional population currents can therefore be written
directly in terms of the prescribed thermal populations and the
instantaneous change of the quasiparticle energy as
\begin{align}
\mathcal{I}_{10,k}(t)
&=
\left[
-2\beta\lambda_{0,k}(t)\lambda_{1,k}(t)
\dot{\epsilon}_k(t)
\right]_{+}
+Q_k(t),
\\[4pt]
\mathcal{I}_{01,k}(t)
&=
\left[
2\beta\lambda_{0,k}(t)\lambda_{1,k}(t)
\dot{\epsilon}_k(t)
\right]_{+}
+Q_k(t).
\end{align}

 Thus, no additional rate construction is required for the TFIM: its instantaneous spectrum determines the thermal population current, while the general formalism of Subsec.~\ref{subsec:engineer} determines the corresponding family of non-negative transition rates. For the numerical results presented below, we primarily consider the
minimal-current realization, $Q_k(t)=0$, such that only the
population-transfer direction required by the instantaneous change of
the Gibbs target is activated.

The performance of the protocol will be assessed by comparing the
dynamically evolved instantaneous populations with their corresponding
thermal target populations. Deviations between these quantities provide
a direct measure of the departure from the prescribed Gibbs trajectory;
the corresponding thermal-excitation measure is introduced explicitly
in Sec.~\ref{subsec:thermal_excitation_deviation}.

\section{Results}
\label{sec:results}
The engineered population-current formalism developed above is now tested numerically against the finite-rate critical crossing introduced in Sec.~\ref{subsec:tfim_driving}, which provides a stringent setting in which to assess its ability to track the prescribed thermal trajectory.

\subsection{Excited-Population Dynamics}
\label{subsec:excited_population}

We first examine the mode-averaged instantaneous excited-state population,
\begin{equation}
    \overline{p}_{e}(t)
    =
    \frac{1}{N_k}
    \sum_{k>0}
    \langle e_k(t)|\rho_k(t)|e_k(t)\rangle ,
    \label{eq:mean_excited_population}
\end{equation}
and compare the engineered dynamics with the evolution induced by the natural thermal reservoir. The corresponding prescribed instantaneous thermal population is
\begin{equation}
    \overline{p}_{e}^{\,\mathrm{th}}(t)
    =
    \frac{1}{N_k}
    \sum_{k>0}\lambda_{1,k}(t).
    \label{eq:mean_thermal_population}
\end{equation}

For the momentum grid defined in Sec.~\ref{sec:tfim}, the smallest positive
momentum is $k_{\min}=\pi/L$. At the critical point $g_c=J$, this gives the
minimum instantaneous level spacing and the associated characteristic
timescale
\begin{equation}
\Delta_{\min}
=
8J\sin\!\left(\frac{\pi}{2L}\right),
\qquad
\tau_{\mathrm{ch}}
=
\frac{1}{\Delta_{\min}}.
\label{eq:finite_size_scale}
\end{equation}
For $L=128$ and $J=1$, these evaluate to
$\Delta_{\min}\simeq9.82\times10^{-2}$ and
$\tau_{\mathrm{ch}}\simeq10.19$.

 Since the duration of the linear ramp from $g_i=2$ to $g_f=0$ is
\begin{equation}
    \tau=\frac{|g_f-g_i|}{v}=\frac{2}{v},
\end{equation}
equating $\tau=\tau_{\mathrm{ch}}$ defines the characteristic velocity
\begin{equation}
    v_{\mathrm{ch}}
    =
    |g_f-g_i|\Delta_{\min}
    \simeq 1.96\times10^{-1}.
    \label{eq:characteristic_velocity}
\end{equation}
%


With these conventions, energies are measured in units of $J$,
times in units of $\hbar/J$, and ramp rates in units of $J^2/\hbar$. The driving rate is reported
relative to the characteristic finite-size scale through the
dimensionless ratio $v/v_{\rm ch}$.
\begin{figure}[t]
    \centering
    \includegraphics[width=\columnwidth]
    {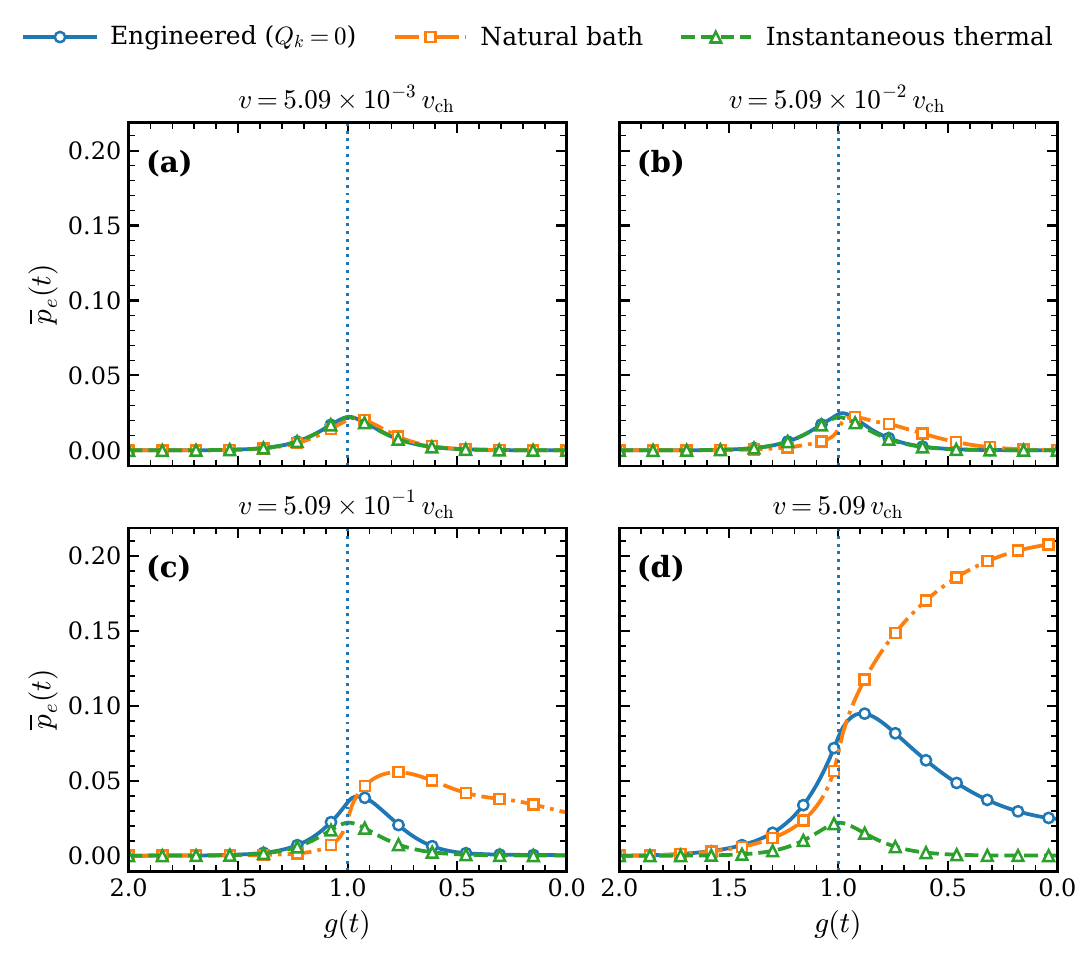}
\caption{Mode-averaged excited-state population for four reduced
driving rates expressed through the ratio $v/v_{\mathrm{ch}}$. The engineered
($Q_k=0$), natural-bath, and instantaneous thermal populations are
shown; the dotted vertical line marks the quantum critical point
$g_c=1$.}
\label{fig:excited_population}
\end{figure}
Fig.~\ref{fig:excited_population} compares the mode-averaged excited-state dynamics for
representative driving rates spanning the regimes below and above the
characteristic scale $v_{\rm ch}$. For $v/v_{\rm ch}<1$, the
engineered dynamics remain close to the instantaneous thermal
population, with deviations increasing as the driving rate approaches
the characteristic scale. At faster driving, $v/v_{\rm ch}>1$, both
dynamical populations depart more substantially from the thermal
target, although the engineered protocol continues to suppress the
excited-state population relative to the natural-bath evolution.

\begin{figure}[t]
    \centering
    \includegraphics[width=\columnwidth]
    {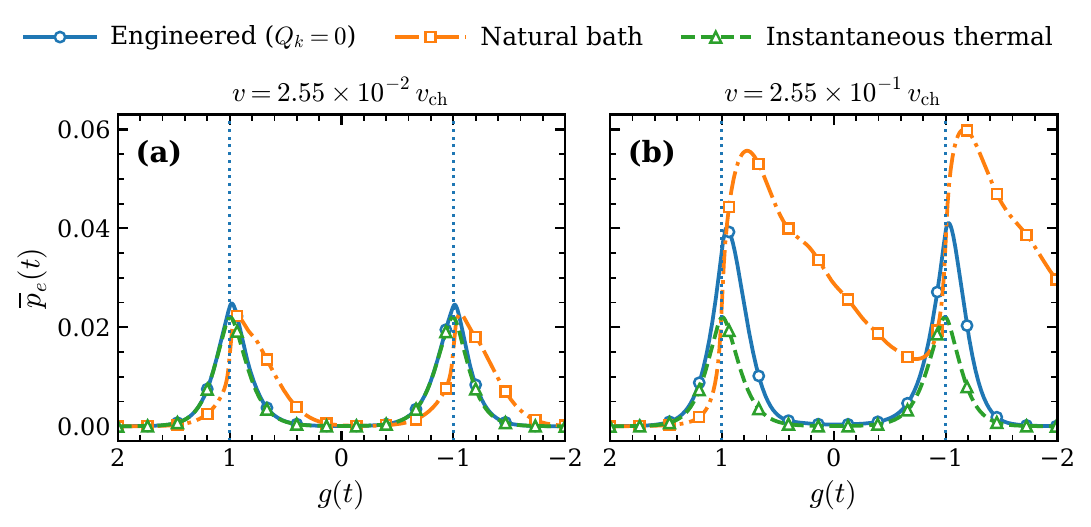}
   \caption{
Mode-averaged excited-state population during a ramp across both
quantum critical points, $g_c=\pm1$. The two panels correspond to
$v/v_{\rm ch}=2.55\times10^{-2}$ and $2.55\times10^{-1}$, respectively,
where $v_{\rm ch}$ is evaluated for the $2\rightarrow-2$ protocol.
The engineered ($Q_k=0$), natural-bath, and instantaneous thermal
populations are shown.
}
    \label{fig:excited_population_two_qcp}
\end{figure}

To examine whether this behavior persists beyond a single critical crossing, Fig.~\ref{fig:excited_population_two_qcp} extends the ramp through the complete ferromagnetic region between $g_c=1$ and $g_c=-1$. The excess excitation generated near the first critical point persists throughout this region and affects the subsequent crossing at $g_c=-1$, particularly under the natural-bath dynamics; in contrast, the engineered evolution remains substantially closer to the instantaneous thermal population across both transitions. The departure from the prescribed thermal trajectory is quantified directly in the following subsection.

\subsection{Thermal Excitation Deviation}
\label{subsec:thermal_excitation_deviation}

The qualitative differences observed in Fig.~\ref{fig:excited_population} motivate a quantitative
measure of the departure from the prescribed instantaneous thermal
trajectory. For the $k$th momentum sector, we denote the actual
instantaneous excited-state population by
\begin{equation}
p_{e,k}(t)
=
\langle e_k(t)|\rho_k(t)|e_k(t)\rangle,
\end{equation}
whereas $\lambda_{1,k}(t)$ denotes the corresponding prescribed thermal
excited-state population.

We therefore define the mode-averaged thermal excitation deviation as
\begin{equation}
\delta n_{\mathrm{ex}}^{\mathrm{th}}(t)
=
\frac{1}{N_k}
\sum_{k>0}
\left|
p_{e,k}(t)-\lambda_{1,k}(t)
\right|.
\label{eq:thermal_excitation_deviation_results}
\end{equation}
This quantity provides a natural finite-temperature extension of the conventional excitation or defect density. In the zero-temperature limit, where the instantaneous thermal excited-state occupation vanishes, $\lambda_{1,k}(t)\rightarrow0$, Eq.~\eqref{eq:thermal_excitation_deviation_results} reduces to
\begin{equation}
    \delta n_{\mathrm{ex}}^{\mathrm{th}}(t)
    \longrightarrow
    \frac{1}{N_k}
    \sum_{k>0}
    p_{e,k}(t)
    \equiv
    n_{\mathrm{ex}}(t).
    \label{eq:thermal_deviation_zero_temperature}
\end{equation}
At finite temperature, however, the excited-state population itself is not an appropriate measure of nonequilibrium excitation, since the target Gibbs state already contains a finite thermal occupation; the deviation from the corresponding instantaneous thermal population therefore provides the relevant measure of excess excitation and will also serve as the quantity used in the subsequent scaling analysis.

\begin{figure*}[ht]
    \centering
    \includegraphics[width=0.94\textwidth]
    {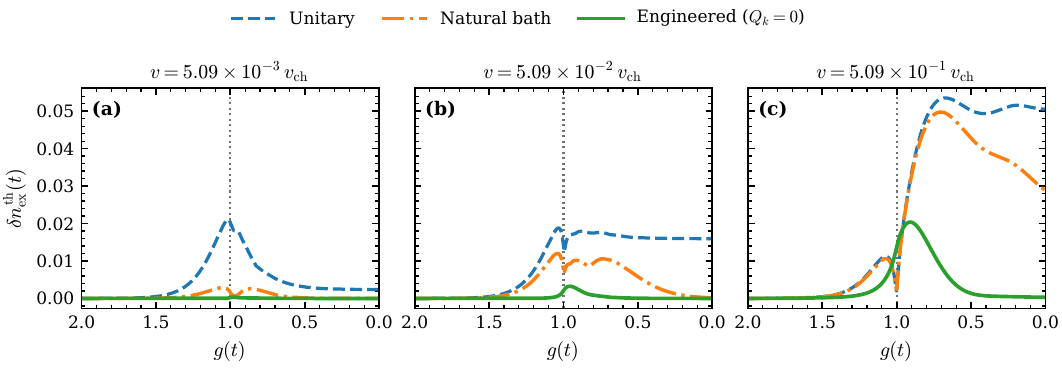}
   \caption{
Thermal excitation deviation
$\delta n_{\mathrm{ex}}^{\mathrm{th}}(t)$ for unitary,
natural-bath, and engineered ($Q_k=0$) dynamics during the
primary ramp $g_i=2\rightarrow g_f=0$.
Panels (a)--(c) correspond respectively to
$v/v_{\rm ch}=5.09\times10^{-3}$,
$5.09\times10^{-2}$, and
$5.09\times10^{-1}$.
These representative rates lie below the characteristic finite-size
driving scale $v_{\rm ch}$ and illustrate the increasing departure
from the instantaneous thermal target as the driving rate is increased.
The dotted vertical line marks the quantum critical point $g_c=1$.
}
    \label{fig:thermal_excitation_deviation}
\end{figure*}

Fig.~\ref{fig:thermal_excitation_deviation} shows that the deviation from the instantaneous thermal target increases with the ramp velocity for all three dynamical cases. While unitary and natural-bath evolution develop pronounced and persistent post-critical deviations, the engineered protocol strongly suppresses the excess excitation across the entire range of velocities considered. For the faster ramps, a finite deviation is generated around the critical region, but under the engineered dynamics it is subsequently driven back toward the prescribed trajectory, in contrast to the residual deviation retained by the unitary and natural-bath evolutions.

\subsection{Suppression of Dynamically Generated Coherence}
\label{subsec:coherence_suppression}
Beyond population tracking, the engineered protocol also shapes the coherence generated during the drive. The prescribed thermal target is diagonal in the instantaneous energy basis, whereas finite-time driving can generate off-diagonal contributions in the actual state $\rho_k(t)$; we therefore examine the mode-averaged coherence
\begin{equation}
    \overline{C}(t)
    =
    \frac{1}{N_k}
    \sum_{k>0}
    \left|
        \rho_{ge,k}^{(E)}(t)
    \right|,
    \label{eq:mean_coherence}
\end{equation}
where $\rho_{ge,k}^{(E)}(t)$ denotes the off-diagonal matrix element of $\rho_k(t)$ in the instantaneous energy eigenbasis.

The general current decomposition introduced in Subsec.~\ref{subsec:engineer} provides an additional degree of freedom through the common population current $Q_k(t)$. Although $Q_k(t)$ does not alter the prescribed net population transfer, it strengthens both the excitation and relaxation channels and therefore enhances dissipative coherence damping. This enhanced bidirectional transition rate directly enters the decay of the off-diagonal density-matrix elements, as seen from the engineered dissipator,
\begin{equation}
    \left(\mathcal{D}^{\mathrm{eng}}_k[\rho_k]\right)_{ij}
    =
    -\frac{W_{01,k}+W_{10,k}}{2}\rho_{ij,k},
    \qquad i\neq j,
    \label{eq:coherence_damping_result}
\end{equation}
the freedom in $Q_k(t)$ can consequently be used to control the dissipative suppression of coherence without modifying the prescribed net population current.

To illustrate this freedom, we consider
\begin{equation}
Q_k(t)=\Gamma\,\lambda_{0,k}(t)\lambda_{1,k}(t),
\end{equation}
where $\Gamma\geq0$ is a rate parameter. With $\hbar=J=1$,
its numerical values are expressed in units of $J/\hbar$.
We consider $\Gamma=1,\,1.5,$ and $2$, together with the
minimal-current choice $Q_k=0$.

\begin{figure*}[t]
    \centering
    \includegraphics[width=0.94\textwidth]
    {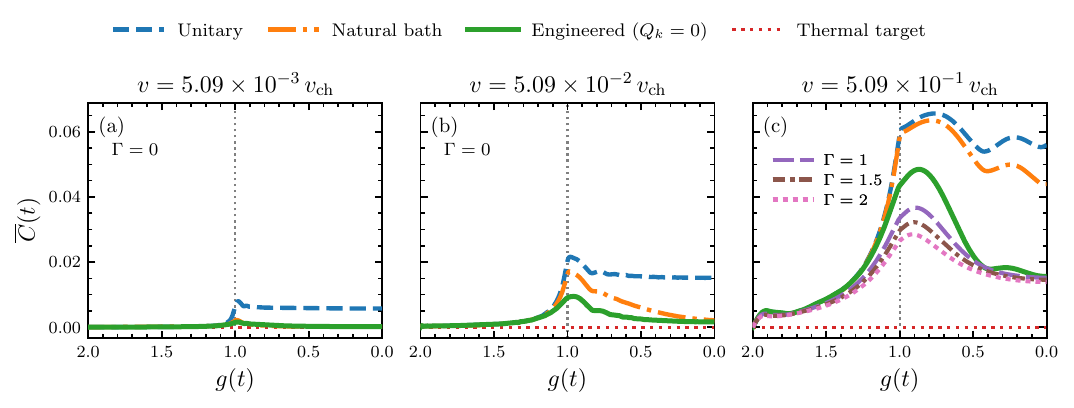}
\caption{
Mode-averaged coherence $\overline{C}(t)$ for the same three driving
rates considered in Fig.~\ref{fig:thermal_excitation_deviation}, comparing unitary, natural-bath, and
engineered dynamics. The minimal-current protocol $Q_k=0$ is shown
together with finite common-current choices
$Q_k=\Gamma\lambda_{0,k}\lambda_{1,k}$ for
$\Gamma=1,\,1.5,$ and $2$ (in units of $J/\hbar$) in panel (c). Increasing $\Gamma$ enhances
coherence damping while leaving the prescribed net population current
unchanged. The instantaneous thermal target has zero coherence in the
instantaneous energy basis. The dotted vertical line marks the quantum
critical point $g_c=1$.
}
    \label{fig:coherence_suppression}
\end{figure*}

As shown in Fig.~\ref{fig:coherence_suppression}, coherence generated during finite-time driving becomes increasingly pronounced with the ramp velocity. The minimal-current protocol $Q_k=0$ already suppresses the coherence relative to the unitary and natural-bath dynamics, while finite $Q_k$ produces a further systematic reduction as $\Gamma$ is increased. Thus, the freedom in $Q_k(t)$ provides an additional control over coherence damping while preserving the same prescribed net population evolution.

\subsection{Scaling of the Residual Thermal Excitation Deviation}
\label{subsec:thermal_scaling}

\begin{figure}[t]
    \centering
    \includegraphics[width=\columnwidth]
    {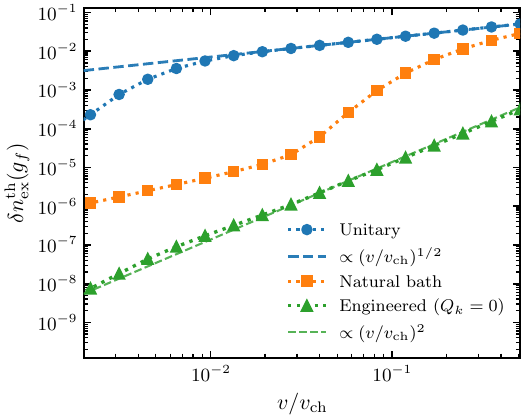}
   \caption{
Final thermal excitation deviation
$\delta n_{\mathrm{ex}}^{\mathrm{th}}(t_f)$ as a function of the
reduced driving rate $v/v_{\mathrm{ch}}$ for unitary, natural-bath,
and engineered ($Q_k=0$) dynamics. Dashed lines indicate the
reference scalings $(v/v_{\mathrm{ch}})^{1/2}$ and
$(v/v_{\mathrm{ch}})^2$ and are not fits to the numerical data.
The unitary dynamics follow the conventional Kibble--Zurek behavior
over the intermediate-rate regime before crossing over at sufficiently
slow driving, whereas the engineered protocol exhibits a substantially
suppressed residual deviation approaching quadratic slow-driving
behavior.
}
    \label{fig:thermal_scaling}
\end{figure}

The time-resolved results above raise a natural question: how does the
residual departure from the instantaneous thermal target scale with the
driving rate? For each protocol, we evaluate the thermal excitation
deviation at the end of the ramp, $g=g_f$, after crossing the quantum
critical point $g_c=1$,
\begin{equation}
\delta n_{\mathrm{ex}}^{\mathrm{th}}(t_f)
=
\frac{1}{N_k}
\sum_{k>0}
\left|
p_{e,k}(t_f)-\lambda_{1,k}(t_f)
\right|.
\label{eq:final_thermal_deviation}
\end{equation}
Fig.~\ref{fig:thermal_scaling} shows this quantity as a function of the
dimensionless driving rate $v/v_{\mathrm{ch}}$. The dashed lines are
not fits to the numerical data; rather, they indicate the reference
power laws $(v/v_{\mathrm{ch}})^{1/2}$ and
$(v/v_{\mathrm{ch}})^2$ for comparison.

For the unitary dynamics, an intermediate-rate regime follows the
characteristic Kibble--Zurek scaling of the one-dimensional
transverse-field Ising model,
\begin{equation}
\delta n_{\mathrm{ex}}^{\mathrm{th}}(t_f)
\sim
\left(\frac{v}{v_{\mathrm{ch}}}\right)^{1/2}.
\label{eq:unitary_kz}
\end{equation}
This behavior reflects the accumulation of nonadiabatic excitations
from modes within the Kibble--Zurek momentum window during passage
through the critical region. At sufficiently small driving rates, however, the unitary dynamics
depart from the thermodynamic Kibble--Zurek scaling and enter a
finite-size slow-driving regime~\cite{sandeep2026supressing}. The $v^{1/2}$ behavior
relies on contributions from a continuum of low-momentum modes within
the Kibble--Zurek momentum window. For a finite chain, the allowed
momenta are discrete and bounded from below by a nonzero minimum
momentum $k_{\min}$. As the driving rate is reduced, the characteristic
momentum window narrows and the excitation density becomes increasingly
dominated by the lowest accessible momentum modes. The dynamics then
resolve the finite-size avoided crossing associated with these modes,
and the continuum approximation underlying the thermodynamic $v^{1/2}$
scaling ceases to apply. Consequently, the unitary dynamics cross over
from the thermodynamic Kibble--Zurek regime to a finite-size
slow-driving regime as $v\rightarrow 0$.

The engineered dynamics display a qualitatively different dependence
on the ramp rate. For the minimal-current construction, $Q_k=0$, the
residual thermal excitation deviation is strongly suppressed relative
to both the unitary and natural-bath dynamics and closely follows the
quadratic reference behavior
\begin{equation}
\delta n_{\mathrm{ex}}^{\mathrm{th}}(t_f)
\sim
\left(\frac{v}{v_{\mathrm{ch}}}\right)^2
\label{eq:engineered_scaling}
\end{equation}
over a broad range of driving rates. We emphasize that the exponent
two is used here as a reference slow-driving scaling rather than being
extracted from a power-law fit. The numerical data approaching this
quadratic dependence indicate that the engineered population currents
suppress the leading finite-rate departure from the prescribed thermal
trajectory, leaving a substantially smaller residual correction.

The natural thermal bath exhibits neither of these simple power laws
over the complete interval. Instead, its residual deviation shows a
pronounced crossover with the driving rate, reflecting competition
between the externally imposed ramp timescale and the finite
bath-induced relaxation timescale. A fixed thermal reservoir can
therefore reduce excitations in parts of the ramp-rate window, but it
does not adapt its population transfer to the changing instantaneous
thermal target.

Taken together, Fig.~\ref{fig:thermal_scaling} demonstrates two distinct
effects of engineered dissipation. First, it reduces the absolute
thermal excitation deviation by several orders of magnitude compared
with the corresponding unitary dynamics. Second, it changes the
finite-rate dependence of this deviation from the conventional
Kibble--Zurek behavior toward a substantially steeper quadratic
slow-driving dependence. This distinction highlights that the
engineered protocol does not merely provide additional relaxation:
its transition currents are constructed from the prescribed thermal
trajectory and therefore actively reshape the finite-time approach to
that trajectory.

\section{Summary and Conclusions}
\label{sec:inference}
In this work, we developed a Markovian dissipative framework for
finite-time thermal-state tracking based on engineered population
currents with non-negative transition rates. The construction begins
from a prescribed instantaneous Gibbs trajectory and determines
transition currents that reproduce the required instantaneous change
of its populations. Applied to the transverse-field Ising model, the
protocol substantially suppresses deviations from the prescribed
thermal populations during a finite-time ramp across the quantum
critical point compared with both coherent evolution and coupling to
a natural thermal reservoir. The freedom associated with the common
current $Q_k(t)$ further provides control over dissipative activity
and coherence damping without modifying the prescribed net population
transfer.

A central result concerns the rate dependence of the residual thermal
excitation deviation following the critical passage. For unitary
dynamics, the numerical results recover the characteristic
Kibble--Zurek behavior
$\delta n_{\mathrm{ex}}^{\mathrm{th}}\sim
(v/v_{\mathrm{ch}})^{1/2}$ over the intermediate-rate regime. At
sufficiently slow driving, the finite system departs from this
thermodynamic scaling as the nonzero finite-size gap becomes dynamically resolved~\cite{sandeep2026supressing}.
In contrast, the engineered minimal-current protocol ($Q_k=0$)
strongly suppresses the residual deviation and approaches a quadratic
slow-driving dependence,
$\delta n_{\mathrm{ex}}^{\mathrm{th}}\sim
(v/v_{\mathrm{ch}})^2$, over a broad range of ramp rates. The natural-bath dynamics instead
exhibit a crossover-like dependence, reflecting the competition
between the driving and relaxation timescales.

The engineered protocol therefore does more than simply enhance
relaxation. By adapting the dissipative population currents to the
prescribed instantaneous thermal trajectory, it reshapes the
finite-rate response while remaining within a time-local Markovian
GKSL description. The accompanying suppression of dynamically
generated coherence and the persistence of improved thermal tracking
for ramps traversing both critical points further demonstrate the
robustness of the construction within the quadratic TFIM setting.

From an experimental perspective, recent developments in reservoir
engineering provide several of the ingredients required for realizing
controlled dissipative dynamics of this kind. In superconducting
circuits, parametrically driven couplings between qubits and lossy
resonator modes have enabled programmable, energy-selective
incoherent pumping and loss, demonstrating direct control over
engineered transition channels~\cite{Guo2026,Shankar2013}.
Complementary progress in trapped-ion platforms has demonstrated
engineered thermal reservoirs with independently tunable temperatures
and dissipation rates, enabling controlled finite-temperature
open-system dynamics~\cite{So2026}. These developments suggest a
possible route toward realizing the population-current construction
considered here, in which the prescribed rates $W_{ij,k}(t)$ could,
in principle, be mapped onto externally controlled system--reservoir
couplings and modulated according to the desired thermal trajectory.

A direct implementation in a many-body spin system nevertheless
requires addressing additional constraints. In particular, the
present construction is formulated in terms of transition channels
resolved in the instantaneous momentum-sector eigenbasis, whose
realization may correspond to nonlocal or otherwise experimentally
nontrivial controls in the underlying spin representation. The
accessible magnitude and bandwidth of the engineered rates, reservoir
response times, and imperfections in the implemented jump operators
will also constrain physical realizations. These considerations
motivate studying the protocol under explicit experimental resource
constraints rather than assuming arbitrary dissipative control.

Future work may therefore investigate the robustness of the observed
scaling with system size and temperature, optimize the common current
$Q_k(t)$ subject to bounds on available dissipative resources, and
extend the construction beyond independent quadratic momentum sectors
to interacting many-body systems. Establishing a platform-specific
mapping between the required time-dependent transition rates and
experimentally accessible reservoir-engineering controls constitutes
an important next step toward translating the present framework into
a realizable dissipative state-preparation protocol.


\bibliography{references}

\end{document}